\documentstyle[prd,aps,preprint]{revtex}

\begin{document}
\draft
\tightenlines
\title{
  \begin{flushright} \small \rm
  hep-ph/9808254 \\
  TIT-HEP-397 \\ 
  KEK-TH-582 \\ 
\end{flushright}
The Singular Seesaw Mechanism 
with Hierarchical Dirac Neutrino Mass}
\author{Yuichi Chikira\footnote{
\tt e-mail: ychikira@th.phys.titech.ac.jp}}
\address{Department of Physics, Tokyo Institute of Technology \\
Oh-okayama, Meguro, Tokyo 152-0033, Japan}
\author{Naoyuki Haba\footnote{
\tt e-mail: haba@phen.mie-u.ac.jp}}
\address{Faculty of Engineering, Mie University, \\
Mie, 514-0008, Japan}
\author{Yukihiro Mimura\footnote{
\tt e-mail: mimura@theory.kek.jp}}
\address{Theory Group, KEK, Oho 1-1, Tsukuba, Ibaraki 305-0801, Japan}
\date{\today}
\maketitle
\begin{abstract}
The singular seesaw mechanism can naturally explain 
the atmospheric neutrino deficit 
by the maximal oscillation between 
$\nu_{\mu_L}$ and $\nu_{\mu_R}$. 
This mechanism can also induce 
three different scales
of neutrino mass squared differences,
which can explain the neutrino deficits of 
three independent experiments 
(solar, atmospheric, and LSND) by
neutrino oscillations. 
In this paper we show that 
 the realistic mixing angles 
 among neutrinos can be obtained
 by introducing the hierarchy
 in the Dirac neutrino mass. 
In the case where Majorana neutrino mass matrix
has rank 2, the solar neutrino deficit is explained
by the vacuum oscillation between $\nu_e$ and $\nu_\tau$.
We also consider the case where Majorana neutrino mass 
matrix has rank 1.
In this case, the mater enhanced 
Mikheyev-Smirnov-Wolfenstein solar neutrino solution is 
prefered as the solution of the solar neutrino deficit.
\end{abstract}
\hspace{1.7 cm} 4.60.Pq, 14.60.St

%
%
\section{Introduction}

According to the recent Super-Kamiokande 
experiment \cite{superk},
 their atmospheric neutrino data 
 indicate the 
 oscillation between $\nu_\mu$ 
 and $\nu_\tau$ or sterile neutrinos 
 with the maximal mixing 
\begin{equation}
\sin^2 2\theta_{\mu x} \sim 1,
\end{equation}
where $x$ represents $\tau$ or 
 sterile neutrinos.
The neutrino mass squared difference $\Delta m^2_{\rm atm}$ is 
 of order  
 $10^{-3} [{\rm eV}^2]$.
 
It is well-known that
 other two independent experiments also 
 imply neutrino oscillations.
One is the solar neutrino experiment.
This experiment implies the oscillation between
 $\nu_e$ and other neutrinos, and there are 
 three possible solutions,
namely, large or small mixing angle 
Mikheyev-Smirnov-Wolfenstein (MSW) 
solution \cite{MSW}, and the 
 vacuum oscillation solution \cite{vac}.
The small (large) angle MSW solution 
 suggests \cite{solar}
\begin{equation}
\sin^2 2\theta_{e x} \sim 10^{-2} \;\; (0.4 {\rm -} 1)
\end{equation}
with mass squared difference 
 of order $ 10^{-5} [{\rm eV}^2]$,
and 
 the vacuum oscillation solution suggests 
\begin{equation}
\sin^2 2\theta_{e x} \sim 0.75{\rm -}1
\label{vacuum-mixing}
\end{equation}
with 
 the mass squared difference of order 
 $ 10^{-10} [{\rm eV}^2]$.
The vacuum oscillation solution is now
 the most suitable solution 
 from the electron energy spectrum 
 of the recent Super-Kamiokande experimental data \cite{Suzuki},
 although the small angle MSW solution has been regarded 
 as the most realistic candidate.
On the other hand, 
 LSND experiment measures oscillation between $\bar\nu_\mu$ 
and $\bar \nu_e$ \cite{LSND}
with a short base line experiment.
Although the 
 confirmation of the LSND results still awaits future
experiments,\footnote{ 
Recent measurements in the KARMEN
detector exclude part of the LSND allowed 
 region \cite{KARMEN}.}
their results indicate the small 
 mixing angle  
\begin{equation}
\sin^2 2\theta_{\mu e} \sim 10^{-2}
\end{equation}
with 
 mass squared difference  
 $\Delta m^2_{\rm LSND} \sim 1 [{\rm eV}^2]$.

The most interesting mechanism which can naturally 
 explain the smallness of neutrino masses is the 
 so-called seesaw mechanism \cite{seesaw}. 
The general 
 mass matrix of neutrinos above $SU(2)_L$ breaking is 
 given by  
\begin{equation}
{\cal M} = \left( \begin{array}{cc} 0 & m_D \\ m_D^T & M_R \end{array} \right),
\end{equation}
 where $m_D$ and $M_R$ represent Dirac and Majorana 
 $3 \times 3$ flavor space mass matrices, 
 respectively.
In the case of $m_D \ll M_R$, 
 there appear three light neutrinos with mass matrix 
\begin{equation}
{\cal M}_{light} = - m_D M_R^{-1} m_D^T .
\end{equation}
This is the essence of the seesaw mechanism.
It is worth noting that
 here it is assumed that there exists 
 inverse matrix of $M_R^{-1}$, that is, 
 $\det M_R \neq 0$. 
The singular seesaw mechanism \cite{singular2,singular1},
 which is also called ``partially broken seesaw
 mechanism'' \cite{FY}\footnote{In this paper
we call this mechanism ``the singular seesaw mechanism''.},
 is just the case of $\det M_R = 0$. 
Then, some light right-handed neutrinos 
 are not integrated out, 
 and behave as sterile neutrinos.
It turns out that mixings between the survived sterile neutrinos 
 and active neutrinos are large in general because of 
 the pseudo-Dirac texture \cite{pD}.
We can use this mechanism to explain 
 the large mixing of the atmospheric neutrino 
 experiment.
%
%
%
If nature adopts four (or more) neutrino oscillations,
 the singular seesaw mechanism supplies one of the most attractive models.

The authors of Ref.\cite{singular2} discussed this singular seesaw 
 mechanism 
 in the case that 
 there is no hierarchy in the Dirac mass matrix  
 $m_D$ and Majorana mass matrix $M_R$. 
They did not take 
 the small mixing of the LSND into account.
In this paper, we study the singular seesaw mechanism 
 by introducing the 
 hierarchy in the Dirac mass matrix $m_D$ 
 in order to explain the small mixing
 of the LSND experiment.
%
We will also study whether  
 the hierarchical Dirac mass can induce 
 not only the small mixing of the LSND experiment
 but also the small mixing of the MSW
 solar neutrino solution.

\par
This paper is organized as follows:
In section II, we will review 
the singular seesaw mechanism briefly. 
In section III, we introduce hierarchical Dirac mass matrix, and
determine the order of parameters.
We show that the vacuum oscillation solution is prefered as the
solution of the solar neutrino deficit in the case where 
Majorana neutrino mass matrix has rank 2 and that the MSW solution
is prefered in the case where Majorana neutrino mass matrix has rank 1.
In section IV, we give summary and discussions.

%
%
\section{Singular Seesaw Mechanism}

At first, we explain the pseudo-Dirac mass 
 texture \cite{pD}.
In the one generation 
 the neutrino mass term 
 above $SU(2)_L$ breaking is
 given by 
\begin{equation}
\label{massmatrix1}
-{\cal L} = \frac{1}{2}\left( \begin{array}{cc}\nu & \nu^C\end{array}\right)
\left( \begin{array}{cc}0 & m \\ m & M \end{array}\right) 
\left( \begin{array}{c} \nu \\ \nu^C \end{array} \right),
\end{equation}
 where $\nu$ and $\nu^C$ represent 
 (two component) left- and right-handed neutrinos, respectively. 
Here we consider the case of 
 $M \ll m$. 
In this case 
 the mass matrix (\ref{massmatrix1}) realizes 
 large mixing angle of 
 $\sin^2 2\theta = \frac{m^2}{m^2 + M^2/4} \sim 1$ between $\nu$ and $\nu^C$. 
The eigenvalues of this mass matrix are $\pm m + M/2$, and
 the neutrino mass squared difference is $\Delta m^2 = 2mM$~\cite{pD}.
This mass term is almost Dirac but not exact, 
 so it is called 
 pseudo-Dirac texture, which can naturally 
 induce the maximal mixing. 
The mass term in 
 the opposite case of $M \gg m$ is that of   
 ordinary seesaw mechanism.

Now let us take three generations into consideration.
We take $m$ and $M$ as 
$3 \times 3$ matrices $m_D$ and $M_M$, respectively in 
 Eq.(\ref{massmatrix1}).
The right-handed Majorana neutrino mass matrix $M_M$ 
is assumed to be rank $2$ (or $1$).
In this case two (one) neutrinos 
 become light by the ordinary seesaw mechanism, 
 and remaining one (two) neutrino has the pseudo-Dirac 
 mass texture. 
{}For example, in the rank-2 case, 
 we can obtain 
 the eigenvalues of four light neutrinos \cite{singular2,singular1} as
\begin{equation}
\beta m, \quad \beta m, \quad{\rm and}\: \pm m + \beta m,
\end{equation}
 where $\beta = m/M$, in the case of no hierarchy in the mass matrices
 $m_D$ and $M_M$.
It is interesting that the two lighter neutrinos' masses
 and the mass splitting for the pseudo-Dirac neutrinos
 are the same scale.
Then three mass squared differences 
 form geometric series as
\begin{equation}
\Delta m^2 = \beta^2 m^2, \quad \beta m^2, \quad {\rm and}\:\: m^2,
\end{equation}
and are favorable to 
 explain three known neutrino oscillation modes,
 namely, 
 solar neutrinos(MSW solution), atmospheric neutrinos and LSND
 \cite{singular1}. 
Furthermore, since the middle scale of the mass squared difference
 for atmospheric neutrinos corresponds to
 pseudo-Dirac neutrinos, 
 its maximal mixing realizes naturally.

However, they can not explain   
 neither the small mixing angle of LSND 
 nor the small angle MSW solution 
 if there is no mass hierarchy in $m_D$ and $M_M$.
We will see that the hierarchical Dirac mass matrices
 lead to different series of mass squared difference,
  which
 is suitable for vacuum oscillation solution for 
 solar neutrino deficit rather than MSW solutions.

The flaws of the singular seesaw mechanism is 
 that the Dirac neutrino mass $m$ needs to be too small (about 1[eV]),
 and it fails the motivation of original seesaw mechanism.
When we incorporate the singular seesaw mechanism into 
 phenomenological models, 
 we need some extra mechanism to apply the small Dirac neutrino mass.
Its smallness will be realized,
 for example, by 
 the non-renormalizable interactions \cite{Langacker},
 though we do not mention the detail in this paper.

%
%
\section{Singular Seesaw Mechanism with 
Hierarchical Dirac Neutrino Mass Matrix}

We introduce the hierarchy 
 in the Dirac neutrino mass matrix as
\begin{equation}
\label{mD}
m_D = \left( \begin{array}{ccc} 
\epsilon' m_{11} & \epsilon' m_{12} & \epsilon' m_{13} \\
\epsilon  m_{21} & \epsilon  m_{22} & \epsilon  m_{23} \\
          m_{31} &           m_{32} &           m_{33}
\end{array} \right) . 
\end{equation}
%
%
We can take the hierarchical parameter $\epsilon$ and
$\epsilon'$ as $\epsilon' \le \epsilon < 1$,
when we do not order the left-handed indices with 
naming of neutrino flavors.
%
In this paper, we do not mention the hierarchical structure
 with respect to right-handed indices.
The mass term of neutrinos is given by 
\begin{equation}
-{\cal L} = m_{Dij}\nu_i\nu_j^C + \frac{1}{2} 
M_{Mij}\nu_i^C\nu_j^C .
\end{equation}
We analyze this model in two cases, 
where the rank of Majorana mass matrix $M_M$ is 1 or 2.

At first, we study the case 
where Majorana mass $M_M$ has rank 2 as 
 $M_M={\rm diag}(M_1,M_2,0)$.
After integrating out 
 the heavy neutrinos\footnote{
 This heavy neutrino mass $M$ turns out to be 1[keV]-1[MeV].
 We integrate out the heavy neutrinos here
 though the scale is lower than the momentum scales of neutrino
 experiments (e.g., about 1[GeV] for atmospheric neutrinos).
 This integrating out method is used
 simply because we display our results clearly.
 Since the heavy neutrinos have very small mixing with
 light neutrinos,
 there is considerable validity to our results below.
}, 
light neutrinos have masses as 
\begin{equation}
-{\cal L} = -\frac{1}{2}\left( \frac{m_{Di1} m_{Dj1}}{M_{1}} + 
     \frac{m_{Di2} m_{Dj2}}{M_{2}} \right) \nu_i \nu_j + m_{Di3}\nu_i\nu_3^C.
\end{equation}
The mass matrix for $(\nu_1,\nu_2,\nu_3,\nu_3^C) \equiv 
(\alpha,\beta,\gamma,s)$ is given by 
\begin{equation}
{\cal M} \sim \left( \begin{array}{cccc} 
-\epsilon'^2 \beta & -\epsilon \epsilon' \beta & -\epsilon' \beta & \epsilon'\\
-\epsilon \epsilon' \beta & -\epsilon^2 \beta  & -\epsilon \beta  & \epsilon \\
-\epsilon' \beta          & -\epsilon \beta    & -\beta           & 1        \\
\epsilon'                 & \epsilon           & 1                & 0        
\end{array}\right) m.
\end{equation}
This matrix is diagonalized as  
\begin{equation}
U^\dagger {\cal M} U \sim
{\rm diag} (\epsilon'^2 \beta m,\ \epsilon^2 \beta m,\
             (1-\beta) m,\ -(1+\beta)m ),
\label{rank-2:diag}
\end{equation}
where
\begin{equation}
U \sim \left( \begin{array}{cccc} 
1 & -{\epsilon'}/{\epsilon} & \epsilon' & -\epsilon' \\
{\epsilon'}/{\epsilon} & 1 & \epsilon & -\epsilon \\
-\epsilon' & -\epsilon & 1 & -1 \\
\epsilon' \beta & -\epsilon \beta & 1 & 1
\end{array} \right).
\end{equation}
Now we estimate the 
 probability of the neutrino oscillation,
 which is given by 
\begin{equation}
P(\nu_\alpha \rightarrow \nu_\beta) = \delta_{\alpha \beta} - 
       4 \sum_{i<j} U_{\alpha i} U^*_{\beta i} U^*_{\alpha j} U_{\beta j} 
       \sin^2 \frac{\Delta m_{ij}^2}{4E}L ,
\end{equation}
where we neglect $CP$ phase for simplicity. 
The oscillation amplitude between $\alpha$ and $\beta$ is given by
\begin{equation}
-4 \sum_{i<j} U_{\alpha i}U^*_{\beta i} U^*_{\alpha j} U_{\beta j}.
\end{equation}
{}From Eq.(\ref{rank-2:diag}),
 we can obtain three scales of mass squared differences
 of 
$\Delta m^2_{12} \sim \epsilon^4 \beta^2 m^2$,
$\Delta m^2_{34} \sim \beta m^2$, and 
$\Delta m^2_{13} \sim \Delta m^2_{14} \sim \Delta m^2_{23} \sim \Delta m^2_{24}
\sim m^2$.
We list the 
amplitudes corresponding to 
these three oscillation in Table I.
The oscillation between $\gamma \leftrightarrow s$
gives a large mixing, which is expected to 
correspond to atmospheric neutrino oscillation. 
Then, we fix
\begin{equation}
\beta m^2 \sim 10^{-3} {\rm [eV^2]}.
\end{equation}
The oscillation with $\Delta m^2 \sim m^2$ may 
 correspond to LSND data, so 
 we fix 
\begin{equation}
m^2 \sim 1 {\rm [eV^2]}.
\end{equation}
Then there remain 
 two patterns whether $(\alpha, \beta)$ 
is assigned as $(e, \tau)$ or $( \tau , e)$. 
Let us consider both possibilities here. 
\begin{description}
\item[(1-1)]
In the case of $( \alpha, \beta) = (e, \tau)$, 
 $\epsilon'$ must be of order 
 $10^{-1}$ from the 
 small mixing of LSND data. 
Then the oscillation with 
 $\Delta m^2 \sim \epsilon^4 \beta^2 m^2 
 \sim 10^{-6} \epsilon^4 {\rm [eV^2]}$ should 
 correspond to 
 the solar neutrino oscillation. 
$(i)$: For the mass squared difference of MSW solution, 
 we must choose the parameter $\epsilon$ to be close to 1.
Then, it turns out that the $\nu_\mu$-$\nu_\tau$ mixing 
 is large with $\Delta m^2 \sim 1[{\rm eV}^2]$.
This oscillation leads to contradiction with 
 the atmospheric neutrino data.
Therefore we can not obtain 
 the MSW solution in this pattern. 
$(ii)$: On the other hand, 
 the vacuum oscillation solution can be realized
 when $\epsilon = O(10^{-1})$. 
We can realize the large mixing of Eq.(\ref{vacuum-mixing})
because the corresponding mixing angle is 
 of order $(\epsilon'/\epsilon)^2$.

\item[(1-2)]
In the case of 
 $(\alpha, \beta) = (\tau, e)$, $\epsilon$
 must be of order $10^{-1}$ from 
 the small mixing of LSND data. 
In this pattern, the mass squared difference corresponding to 
the solar neutrino oscillation should be 
 of order $10^{-10}$[eV$^2$]. 
Therefore, only vacuum oscillation solution can be allowed.
To explain the large mixing of the solution, 
$\epsilon'$ must be satisfy $\epsilon \sim 10^{-1}$.
This is the same parameters as the case of $(ii)$ in 
 {\bf (1-1)}.
\end{description}

Next, let us consider 
the case where Majorana mass $M_M$ has rank 1 as 
 $M_M = {\rm diag}(M,0,0)$.
The mass term of neutrinos is given by 
\begin{equation}
-{\cal L} = m_{Dij} \nu_i \nu_j^C + \frac{1}{2}M \nu_1^C \nu_1^C.
\end{equation}
After integrating $\nu_1^C$ out, 
 light neutrinos 
 $(\nu_1, \nu_2, \nu_3, \nu_2^C, \nu_3^C) \equiv 
 (\alpha, \beta, \gamma, s_1, s_2)$ 
 have masses as 
\begin{equation}
{\cal M} \sim \left( \begin{array}{ccccc} 
-\epsilon'^2 \beta & -\epsilon' \epsilon \beta & -\epsilon' \beta 
& \epsilon' & \epsilon' \\
-\epsilon \epsilon' \beta & -\epsilon^2 \beta & -\epsilon \beta &
\epsilon & \epsilon \\
-\epsilon' \beta & -\epsilon \beta & -\beta & 1 & 1 \\
\epsilon' & \epsilon & 1 & 0 & 0 \\
\epsilon' & \epsilon & 1 & 0 & 0 \\ \end{array} \right)m.
\end{equation}
This matrix can be diagonalized as
\begin{equation}
U^\dagger {\cal M} U \sim 
{\rm diag} ( \epsilon'^2 \beta m,\ (\epsilon - \epsilon^2 \beta) m, \ 
            -(\epsilon + \epsilon^2 \beta) m,\ (1 - \beta) m, \ 
            -(1 + \beta) m ),
\end{equation}
where
\begin{equation}
U \sim \left( \begin{array}{ccccc}
1 & -{\epsilon'}/{\epsilon} & -{\epsilon'}/{\epsilon} & 
\epsilon' & \epsilon' \\
{\epsilon'}/{\epsilon} & 1 & 1 & \epsilon & \epsilon \\
\epsilon' & -\epsilon & -\epsilon & 1 & 1 \\
\epsilon' \beta & -1 & 1 & 1 & -1 \\
\epsilon' \beta & 1 & -1 & 1 & -1 
\end{array} \right).
\end{equation}
There are four scales of mass squared differences as  
$\Delta m^2 \sim \epsilon^3 \beta m^2$, $\beta m^2$, 
 $\epsilon^2 m^2$
and $m^2$.
The oscillation amplitudes corresponding to these oscillation modes are 
listed in Table II.
The atmospheric neutrino oscillation can be regarded as 
 $\gamma \leftrightarrow s_1,s_2$.
Therefore, the mass squared difference 
 $\beta m^2$ must be
 of order $10^{-3}$[eV$^2$].
Then, the mass squared difference 
 corresponding to the solar neutrino oscillation
 should be $\epsilon^3 \beta m^2 {\rm [eV^2]} 
 \sim 10^{-3} \epsilon^3 {\rm [eV^2]}$.
There are two candidates for the mass squared differences of LSND,
namely,
$\Delta m^2_{\rm LSND} \sim \epsilon^2 m^2$ or $\Delta m^2_{\rm LSND} \sim
m^2$. 
Here we consider the both possibilities.
\begin{description}
\item[(2-1)]
In the case of $(\alpha, \beta) = (e, \tau)$, 
 $\epsilon'$ must be of order $10^{-1}$ from
 the small mixing of LSND data. 
As for the solar neutrinos,
 the vacuum oscillation solution 
 is excluded because
 the parameter $\epsilon$ can not be 
 smaller than $\epsilon' \sim 10^{-1}$. 
Then 
 we consider the parameter $\epsilon = O(10^{-1})$ 
 in order to obtain the suitable 
 mass squared difference for the 
 MSW solution.
In this case 
 $\epsilon'/\epsilon$ tend to becomes close to $1$, and
in this case, the $\nu_e \leftrightarrow 
 \nu_\tau$ oscillation with $\Delta m^2 \sim 
\epsilon^2 m^2$ becomes large mixing. 
However, the large mixing of the order of $(\epsilon'/\epsilon)^2 \sim 1$
with $\Delta m^2 > 10^{-3} {\rm [eV^2]}$ is excluded from CHOOZ 
experiment \cite{chooz}, and
we should choose the mixing $(\epsilon'/\epsilon)^2$ to be
smaller than $O(10^{-1})$.
This choice of parameters leads the solar neutrino solution to 
the small angle MSW solution.  
Therefore, $(\epsilon'/\epsilon)^2$ appears to be 
 of order $10^{-2}$. 
In order to get such a $(\epsilon'/\epsilon)^2$, we need delicate 
tuning of parameters.
%
\item[(2-2)]
In the case of $(\alpha, \beta) = (\tau, e)$,
 $\epsilon'$($\epsilon$) must be of order $10^{-1}$
 from small mixing amplitude of LSND with  
 $\Delta m^2_{\rm LSND} \sim \epsilon^2 m^2$
 ($\Delta m^2_{\rm LSND} \sim m^2$). 
For the same reason of {\bf (2-1)}, 
the vacuum oscillation solution is excluded,
and 
the parameter $\epsilon$ should be 
 chosen to be of order $10^{-1}$
 for the MSW solution.
Though the large angle MSW solution
 through $\nu_e \leftrightarrow \nu_s$ oscillation mode seems to be possible,
 it is not allowed at the 99\% C.L. \cite{LMA}
 in two flavor analysis.
Therefore, in this case, we can not help but consider other oscillation
mode,
 namely, $\nu_e \leftrightarrow \nu_\tau$,
 as the solution for solar neutrino deficit.
As we mentioned in {\bf (2-1)}, 
the small angle MSW solution through $\nu_e \leftrightarrow \nu_\tau$ oscillation mode seems to be possible.
However, since the mixing of $\nu_e$ and $\nu_s$ is large, 
we need the detail analysis of 
three generation mixing in this case.  


\end{description}

%
%
\section{Conclusion}

The recent atmospheric neutrino data of 
 Super-Kamiokande suggests 
 the maximal mixing between $\nu_{\mu}$ and 
 other neutrinos.
The singular seesaw mechanism is one of the most interesting 
 scenario that can
 naturally explain 
 this large mixing angle between 
 $\nu_{\mu_L}$ and $\nu_{\mu_R}$. 
This mechanism can also induce three independent
 mass 
 squared differences, which are suitable for 
 the solutions of the solar and atmospheric
 neutrino anomalies, and the LSND data.
The original scenario in Ref.\cite{singular2} can not 
 explain neither the small mixing angle of 
the LSND data nor small angle solution of MSW.
Thus, we introduce the hierarchy in the Dirac neutrino mass matrix,
 and reanalyzed the singular seesaw mechanism. 
As the results, we can obtain 
 the small mixing solutions of the 
 LSND and MSW as follows.
In the case of rank-2 Majorana mass, 
the Dirac mass matrix should be the form of 
\begin{equation}
\left( \begin{array}{ccc}
    \epsilon m_{ee} &     \epsilon m_{e\mu} & \epsilon m_{e\tau} \\
          m_{\mu e} &           m_{\mu \mu} &          m_{\mu \tau} \\
\epsilon m_{\tau e} & \epsilon m_{\tau \mu} & \epsilon m_{\tau \tau} \\
\end{array} \right),
\end{equation}
where dimensionless parameter $\epsilon$ is 
 of order $10^{-1}$ and 
 $m_{\alpha \beta}\sim 1$[eV].
The non-zero elements of Majorana mass should be of order 1[keV].
It is important that 
the solar neutrino deficit can be explained by the vacuum oscillation 
between $\nu_e$ and $\nu_\tau$,
in contrast to the original framework of Ref.\cite{singular2}.

In the case of rank-1 Majorana mass, 
 the small angle MSW solution 
 is suitable for 
 the solar neutrino oscillation.
The Dirac mass matrix should be 
\begin{equation}
\left( \begin{array}{ccc}
    \epsilon m_{ee} &     \epsilon m_{e\mu} & \epsilon m_{e\tau} \\
          m_{\mu e} &           m_{\mu \mu} &          m_{\mu \tau} \\
\epsilon' m_{\tau e} & \epsilon' m_{\tau \mu} & \epsilon' m_{\tau \tau}
          \\
\end{array} \right),
\end{equation}   
where $\epsilon$ is of order $10^{-1}$ and $\epsilon'$ should satisfy the
condition $(\epsilon'/\epsilon)^2 < 10^{-1}$.
There is an extra oscillation mode 
$\Delta m^2 \sim 10^{-2} {\rm [eV^2]}$ 
or $\Delta m^2 \sim 10^2 {\rm [eV^2]}$.

\par
 Finally, we would like to comment about
Big Bang nucleosynthesis (BBN) in the sterile
scenario.
The ratio of deuterium to hydrogen and the
abundance of $^4$He are determined by the
ratio of neutrons to protons at the time
when the weak interaction freeze out. 
The effective number of light neutrino flavors
$N_{\nu}$ contribute to the energy density,
which influences the expansion rate.
Thus, we can obtain the upper limit of $N_{\nu}$ from
the BBN constraint \cite{Hata}.
Although the standard BBN scenario shows
 $N_{\nu} \leq 3.6$ \cite{Hata}, 
 the large lepton number
 asymmetry in the early universe may allow
 $N_{\nu}=4$ \cite{Foot}, which corresponds to 
 the rank-2 case in this paper. 
On the other hand, 
 rank-1 case induces $N_{\nu}=5$, 
 since $\nu_{\tau}$-$\nu_s$  
 mixing is large. 
Thus this case needs extra mechanism which suppress
 the effective number of neutrinos in 
 the early universe\footnote{
One possibility is to consider the
 large uncertainty of the systematic error in 
 $^4$He abundance \cite{Gaiser}. 
}.


\acknowledgments

N.H would like to thank Professor S. Raby and
Dr. K. Tobe for helpful discussions.
Y.M. wishes to thank Institute for Cosmic Ray Research, 
University of Tokyo, in which most of this research
was carried out.
This work was partially supported by 
JSPS Research Fellowships for Young Scientists (Y.C. and Y.M.).

%
%

\begin{table}[t]
\begin{center}
\begin{tabular}{cccc}
& $\Delta m^2 \sim \epsilon^4 \beta^2 m^2$ & $\Delta m^2 \sim \beta m^2$ 
& $\Delta m^2 \sim m^2$ \\ \tableline
$\alpha \leftrightarrow \beta$ & 
$\left( \frac{\epsilon'}{\epsilon} \right)^2$ & $\epsilon^2 \epsilon'^2$ &
$\epsilon'^2$ \\ 
$\alpha \leftrightarrow \gamma$ &
$\epsilon'^2$ & $\epsilon'^2$ & $\epsilon'^2$ \\ 
$\alpha \leftrightarrow s$ & 
$\epsilon'^2 \beta^2$ & $\epsilon'^2$ & $\epsilon'^2 \beta$ \\
$\beta \leftrightarrow \gamma$ &
$\epsilon'^2$ & $\epsilon^2$ & $\epsilon^2$ \\ 
$\beta \leftrightarrow s$ &
$\epsilon'^2 \beta^2$ & $\epsilon^2$ & $\epsilon^2 \beta$ \\ 
$\gamma \leftrightarrow s$ &
$\epsilon'^2 \epsilon^2 \beta^2$ & $1$ & $\epsilon^2 \beta$ \\ 
\end{tabular}
\end{center}
\caption{Oscillation amplitudes in the case of rank-2}
\end{table}

\begin{table}[t]
\widetext
\begin{center}
\begin{tabular}{ccccc} 
& $\Delta m^2 \sim \epsilon^3 \beta m^2$ & $\Delta m^2 \sim \beta m^2$ &
$\Delta m^2 \sim \epsilon^2 m^2$ & $\Delta m^2 \sim m^2$ \\ \tableline
$\alpha \leftrightarrow \beta$ & 
$(\frac{\epsilon'}{\epsilon})^2$ & $\epsilon^2 \epsilon'^2$ & 
$(\frac{\epsilon'}{\epsilon})^2$ & $\epsilon'^2$ \\ 
$\alpha \leftrightarrow \gamma$ &
$\epsilon'^2$ & $\epsilon'^2$ & $\epsilon'^2$ & $\epsilon'^2$ \\ 
$\alpha \leftrightarrow s_1,s_2$ &
$(\frac{\epsilon'}{\epsilon})^2$ & $\epsilon'^2$ & 
$\frac{\epsilon'^2}{\epsilon}\beta$ & $\frac{\epsilon'^2}{\epsilon}$ \\ 
$\beta \leftrightarrow \gamma$ &
$\epsilon^2$ & $\epsilon^2$ & $\epsilon'^2$ & $\epsilon^2$ \\ 
$\beta \leftrightarrow s_1,s_2$ &
$1$ & $\epsilon^2$ & $\frac{\epsilon'^2}{\epsilon}\beta$ & $\epsilon$ \\ 
$\gamma \leftrightarrow s_1,s_2$ &
$\epsilon^2$ & $1$ & $\epsilon \epsilon' \beta$ & $\epsilon$ \\
\end{tabular}
\end{center}
\caption{Oscillation amplitudes in the case of rank-1}
\end{table}


\begin{references}
%
\bibitem{superk} Super-Kamiokande Collaboration, Y. Fukuda et al., 
hep-ex/9803006; hep-ex/9805006; hep-ex/9807003;
talk by T. Kajita at {\it Neutrino-98}, Takayama, Japan, June 1998. 
%
\bibitem{MSW}
L. Wolfenstein, Phys. Rev. {\bf D17}, 2369 (1978). \\
S. P. Mikheyev and A. Yu. Smirnov, Yad. Fiz. {\bf 42}, 1441 (1985)
[Sov. J. Nucl. Phys. {\bf 42}, 913 (1985)]; Nuovo Cim. {\bf 9C}, 17 (1986). 
%
\bibitem{vac}
{\it See example,} V. Barger, R. J. N. Phillips, and K. Whisnant, 
Phys. Rev. Lett. {\bf 69}, 3135 (1992).
%
\bibitem{solar}
GALLEX Collaboration, Phys. Lett. {\bf B285}, 390 (1992).\\
P. I. Krastev and S. T. Petcov, Phys. Lett. {\bf B299}, 99 (1993).\\
G. Fiorentini et. al., Phys. Rev. {\bf D49}, 6298 (1994).\\
N. Hata and P. G. Langacker, Phys. Rev. {\bf D56}, 6107 (1997).
%
\bibitem{Suzuki}
Talk by Y. Suzuki at {\it Neutrino-98}, Takayama, Japan, June 1998.
%
\bibitem{LSND}
LSND Collaboration, C. Athanassopoulos et al., 
Phys. Rev. Lett. {\bf 75}, 2650 (1995); $ibid$. 
{\bf 77}, 3082 (1996); nucl-ex/9706006. 
%
\bibitem{KARMEN}
The LSND results will be tested by the KARMEN experiment, 
talk by B. Armbruster at 33rd Rencontres de Moriond :
 Electroweak Interactions and Unified Theories, 
Les Arcs, France, March 1998, and talk by 
B. Zeitnitz at {\it Neutrino 98}, Takayama, Japan, June 1998, 
and also by the BooNE experiment, E. Church et al., nucl-ex/9706011. 
%
\bibitem{seesaw}
T. Yanagida, in: O. Sawada and A. Sugamoto(eds.), {\it Proc. of 
the Workshop on 
the Unified Theory and Baryon Number in the Universe} (KEK report 79-18, 
1979) p.95.\\
M. Gell-Mann, P. Ramond, and R. Slansky, in: P. van Nieuwenhuizen and 
D. Z. Freedman(eds.), {\it Supergravity} (North Holland, Amsterdam, 1979) 
p.315.
%
\bibitem{singular2} E. J. Chun, C. W. Kim, and U. W. Lee, 
Phys. Rev. {\bf D58}, 093003 (1998). 
%
\bibitem{singular1} S. L. Glashow, Phys. Lett. {\bf B256}, 255 (1991).
%
\bibitem{FY} M. Fukugita and T. Yanagida, Phys. Rev. Lett. {\bf 66}, 2705 (1991).
%
\bibitem{pD}
K. Kobayashi, C. S. Lim and M. M. Nojiri, 
Phys. Rev. Lett. {\bf 67}, 1685 (1991). \\
H. Minakata and H. Nunokawa, Phys. Rev. {\bf D45}, 3316 (1992).  
%
\bibitem{Langacker}
P. Langacker,
Phys. Rev. {\bf D58}, 093017 (1998).
%
\bibitem{chooz} CHOOZ Collaboration, M. Apollonio et al.,
Phys. Lett. {\bf B420}, 397 (1998).
%
\bibitem{LMA} J. N. Bahcall, P. I. Krastev, and A. Yu. Smirnov,
hep-ph/9807216.
%
\bibitem{Hata}
See for example, 
N. Hata, G. Steigman, S. Bludman, and P. Langacker,
Phys. Rev. {\bf D55}, 540 (1997).
%
\bibitem{Foot}
See for example,
R. Foot and R. R. Volkas,
Phys. Rev. {\bf D56}, 6653 (1997).
%
%

\bibitem{Gaiser}
A. Geiser, hep-ph/9810493. 

\end{references}
\end{document}